# Fluctuation-enhanced sensing[1]

(Keynote invited talk)

L.B. Kish[(+)], G. Schmera[(++)], Ch. Kwan[(x)], J. Smulko[(xx)], P. Heszler[(*)], C.-G. Granqvist[(**)]

[(+)]Texas A&M University, Department of Electrical and Computer Engineering, College Station, TX 77843-3128, USA

[(++)]Space and Naval Warfare System Center, Signal Exploitation & Information Management, San Diego, CA 92152-5001, USA

[(x)]Signal Processing, Inc., 13619 Valley Oak Circle, Rockville, MD 20850, USA

[(xx)]Gdansk University of Technology, WETiI, ul. G. Narutowicza 11/12, 80-952 Gdansk, Poland

[(*)]University of Szeged, Department of Experimental Physics, Dom ter 9, Szeged, H-6720, Hungary

[(**)]Department of Engineering Sciences, The Ångström Laboratory, Uppsala University, P.O. Box 534, SE-751 21 Uppsala, Sweden

## ABSTRACT

We present a short survey on fluctuation-enhanced gas sensing. We compare some of its main characteristics with those of classical sensing. We address the problem of linear response, information channel capacity, missed alarms and false alarms.

## 1. INTRODUCTION

### 1.1. Fluctuation-enhanced sensing

The classical way of physical and chemical sensing involves the measurement of the value of a physical quantity in the detector/sensor. Recently, a new method has been proposed for chemical and gas sensing and developed that mimics the biological way of sensing where the sensed agent changes the statistics of the neural output, which is a pulse noise. Thus noise carries the sensory information. We call this type of sensing *Fluctuation-Enhanced Sensing* (FES). We note that FES has been used long time ago to measure certain physical quantities under difficult conditions. For example, the measurement of Johnson noise voltage of resistors has been utilized to determine temperature in cryogenic applications for a long time.

In Figure 1, the usual Johnson voltage noise thermometry is compared with classical resistor thermometry. For a resistor thermometer, the $R(T)$ function must be known and a biasing DC current, which is heating the thermometer and causes errors, and precise current and voltage measurements are needed. In the usual Johnson-noise thermometry, the $R(T)$ function is not needed but the resistance value *R* still must be known/determined (measuring it causes heating at classical measurements) and precise current and voltage measurements are needed.

---

[1] Invited papers with similar scientific content and some overlap with the present paper will be presented at the NANO-DDS conference in Washington DC (June 2007), and the ICNF conference in Tokyo (September 2007). A similar literature survey is published as an invited paper in Nanotechnology Perceptions. The present paper is modified and edited to match the community of SPIE's Fluctuations and Noise Symposium.

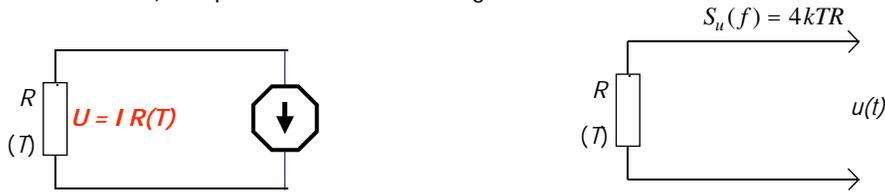

Figure 1. Left: Resistor thermometer; the *R(T)* function must be known and a biasing DC current, which is heating the thermometer and causes error, and precise current and voltage measurements are needed. Right: The usual Johnson-noise thermometry; the *R(T)* function is not needed but the resistance value *R* must be known and precise current and voltage measurements are needed.

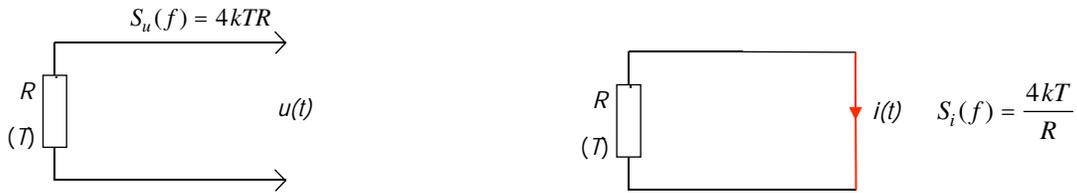

Figure 2. Fluctuation-enhanced absolute thermometry based on Johnson noise and first principles. Left: The Johnson voltage noise is measured. Right: The Johnson current noise is measured. The two measurements provide both the absolute temperature *T* and the resistance *R*. No heating occurs and no preliminary knowledge is needed except precise current and voltage measurements.

In Figure 2, the method utilizes the full power of FES. This absolute thermometry is based on Johnson noise and first principles. The Johnson voltage noise and current noise are measured. They result in a system of two separate equations. Their solution provides both the temperature and the resistance:

$$T = \frac{\sqrt{S_u S_i}}{4kT} \qquad R = \frac{S_u}{S_i} \qquad (1)$$

No heating occurs and only precise current and voltage measurements are needed. This solution also indicates that FES is a powerful tool, it does not mean that it is an easy/cheap measurement.

The focus topic of our paper is *fluctuation-enhanced sensing of gases* which is far less "clean" as first principle measurements. Classical gas sensing methods are many orders of magnitude less sensitive than the nose of dogs or even that of humans. So, how do biological noses do the job? They contain a large array of olfactory neurons which communicate stochastic voltage spikes to the brain. When odor molecules are adsorbed by a number of neurons, the statistical properties of these stochastic spikes change. The brain decodes the changes in statistics and matches the result with an odor database in memory.

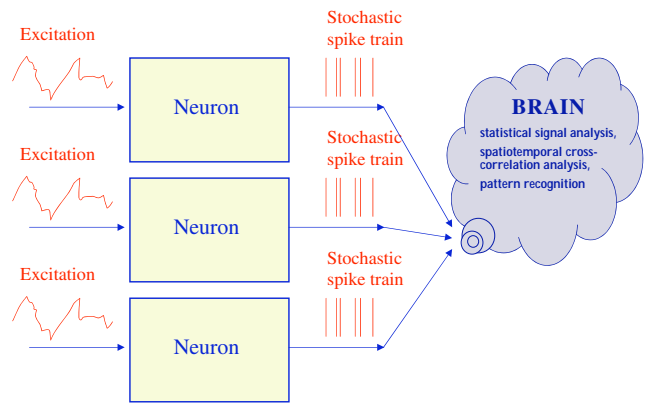

Figure 3. Fluctuation-enhanced odor and taste sensing by humans and animals.

The first step is similar to the classical sensing method: the value of a physical quantity in the sensing medium is measured, for example the output voltage of a chemical sensor. Then the microscopic spontaneous fluctuations of these measurements are strongly amplified (typically 1-100 million) and the statistical properties of these fluctuations are analyzed. These fluctuations are due to the dynamically changing molecular-level interactions between the odor molecules and the sensing media, thus they contain the chemical signature of the odor. The results are compared with a statistical pattern database to identify the odor. The new method has been named "fluctuation-enhanced" sensing.

In the next sections we first address the problem of classical gas sensing and then history of FES. Then we will address some of the many practical problems of FES.

### 1.2. Classical gas sensing

Concerns about *outdoor* air-pollution are widely spread. However, it is less known that serious health-related problems may emerge also from the *indoor* environment. Indoor air contains a wide variety of volatile organic compounds (VOCs, e.g., formaldehyde and vapors of organic solvents), and a number of these VOCs have a higher concentration indoors than outdoors [1]. Exposure to VOCs has been suggested to cause, e.g., mucous irritation, neurotoxic effects (fatigue, lethargy, headache, etc.) and nonspecific reactions (e.g., chest sounds and asthma-like symptoms) [2, 3]. It is clear that precise air quality monitoring is of great importance in both in- and outdoor environments. This requires sensors capable of detecting low concentrations of $CO_2$, $CO$, $SO_2$, $NO_x$, $O_3$, $H_2S$, $HF$, $Cl_2$, VOCs, etc. sensitively and selectively. (The listed gases have been selected as they have toxic effects [4].) This huge need could be best fulfilled with simple, cheap, and replaceable sensors, most preferably electronic, semiconductor type that can be easily integrated into existing monitoring and ventilation systems.

    The operation principle of the "classical", Taguchi-type semiconductor gas sensors is based on the change of the sensor resistance as the gas to be sensed is adsorbed on the sensor surface [5]. This type of sensors represents a low-cost option to the standardized and bulky methods (e.g., gas chromatography or mass spectroscopy). Metal-oxides, e.g., $SnO_2$, $TiO_2$, $ZnO$, $Mn_2O_3$ and $WO_3$, are most commonly used as sensor materials [6]. There is continuous work for improving the sensor performance, including sensitivity and, most importantly, the chemical selectivity of these kinds of sensors.

    Regarding sensitivity, nanotechnology—particularly the use of Nanostructured Materials (NsMs)—offers new possibilities in this area. The characteristic structural length of a NsM is typically 1 to 100 nanometers. One class of NsMs is composed of nanoparticles or nanocrystals, and in a porous structure these materials exhibit high surface area, which can be orders of magnitude higher than that of coarser, micro-grained materials, therefore increasing sensitivity of the gas sensors [7, 8]. It is likely that not only the high surface area but the actual nanostructure (e.g., neck and grain boundary formation between nanograins) also plays a role for sensitivity enhancement of NsMs [9]. Sensitivity can also be improved by doping the oxide materials [6, 10].

    Chemical selectivity of semiconductor gas sensors can be boosted by operating an array of sensors, each of them having different sensitivity for different gases (also referred to as "electronic nose") [11]. This can be achieved by, e.g., using different (or doped) sensor materials or by operating the sensors at different temperatures. The output of sensor arrays is then analyzed by pattern recognition methods [12]. Investigating the dynamic response of temperature-modulated sensors is also a possible way for improving chemical selectivity [13]. However, lack of selectivity is still a significant problem for the widespread use of semiconductor gas sensors.

### 1.3. Short survey of the history of fluctuation-enhanced gas sensing

While some optical chemical sensors analyze the absorption or emission spectrum of gases and therefore are able to generate a pattern, most chemical sensors produce a single number output only. For example the steady-state value of a Taguchi sensor, or the steady-state current value of a MOS sensor, are such signals. To generate a separate pattern corresponding to different chemical compositions, a number (6 to 40) of different types of sensors are needed, which makes the system expensive and unreliable for practical applications. On the other hand Fluctuation-Enhanced Sensing (FES), see Figure 4, is able to generate a complex pattern by the application of a single sensor [14-22].

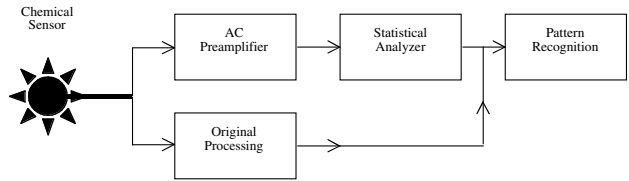

Figure 4. Fluctuation-enhanced chemical sensing.

FES means that, instead of using the mean value (time average) of the sensor signal, the small stochastic fluctuations around the mean value are amplified and statistically analyzed. Due to the grainy structure of resistive film sensors, these materials exhibit significantly (several orders of magnitude) increased electronic resistance fluctuations compared the case of to single crystalline materials, and these fluctuations are strongly influenced by the random walk (diffusion) dynamics of agents in the vicinity of intergrain junctions and by adsorption-desorption noise. Stochastic analytical tools are used to generate a one-dimensional or two-dimensional pattern from the time fluctuations. The analysis of these patterns can be done in the classical way by using pattern recognition tools.

The history of FES goes back more than a decade [14-37]. The name "Fluctuation-Enhanced Sensing" was created by John Audia (SPAWAR, US Navy) in 2001. In this paper we mostly focus on journal papers and neglect the vast body of conference contributions, except in cases where conference papers (or patents) have reserved the priority.

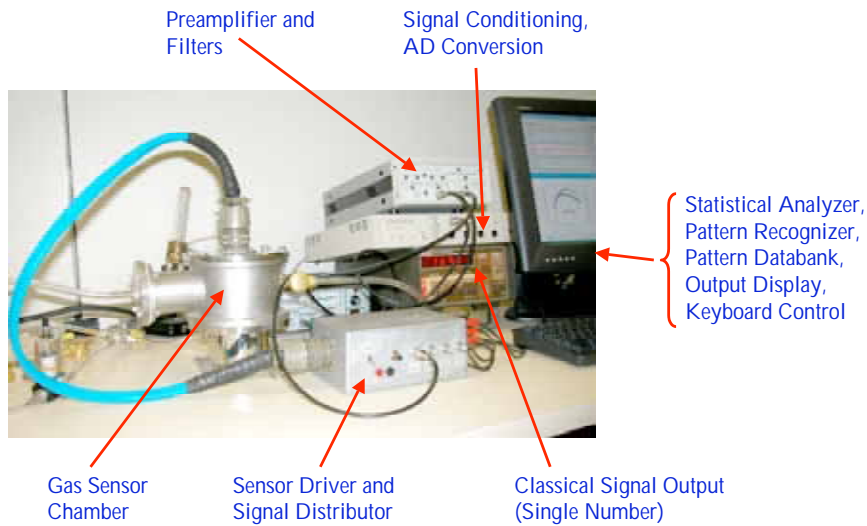

Figure 5. Fluctuation-enhanced gas sensing setup at the Department of Electrical and Computer Engineering at Texas A&M University.

Using electrical noise (spontaneous fluctuations) to identify chemicals was first proposed by Neri and coworkers [14,15] in 1994-95 by showing the sensitivity of conductance noise spectra of conducting polymers as a function of the ambient gas composition. In 1997, Gottwald and coworkers [16] published similar effects for the conductance noise spectrum of semiconductor resistors with non-passivated surfaces. The first mathematical analysis of generic FES systems, with the sensor number requirement versus the number of agents, was done by Kish and coworkers in 1998 [17-19]. The possibility of "freezing the smell" in a Taguchi sensor was first demonstrated by Vajtai [18] and later a more extensive analysis was published by Solis et al [20]. In 2001, Smulko et al were the first to use higher-order statistics to enhance the extracted information from the stochastic signal component [21,26,29]. Hoel et al showed FES via invasion noise effects at room-temperature in nanoparticle films [22]. Schmera and coworkers analyzed the situation of Surface Acoustic Wave (SAW) sensors and predicted the FES spectrum for SAW and MOS sensors with surface diffusion [23-24]. Commercial-On-The-Shelf (COTS) sensors with environmental pollutants and gas combinations were also studied [25,29,30]. In nanoparticle sensors with a temperature gradient, the possibility of using the noise of the thermoelectric voltage for FES was demonstrated [28]. Ederth et al analyzed the sensitivity enhancement in the FES mode and compared it to the classical mode in nanoparticle sensors and found an enhancement of a factor of 300. Gomri et al [32,33] published FES theories for the cases of adsorption-desorption noise and chemisorption-induced noise. Huang et al explored the possibility of using FES in electronic tongues [34].

## 2. ON THE SENSITIVITY AND SELECTIVITY IN FLUCTUATION-ENHANCED SENSING

The statistics of the microscopic fluctuations in a system are rich and sensitive sources of information about the system itself. They are extremely sensitive because the perturbations of microscopic fluctuations require only very small energy. On the other hand, the related statistical distribution functions are data arrays, and thus they can contain orders of magnitude more information then a single number represented by the mean value of the sensor signal used in classical sensing.

The underlying physical mechanism behind the enhanced *sensitivity* is the temporal fluctuations of the agent's or its fragment's concentration at various points of the sensor volume where the sensitivity of the resistivity against the agent is different. This effect will generate stochastic fluctuations of the resistance and the sensor voltage during biasing the sensor with a DC current. The voltage fluctuations can be extracted (by removing the mean value by AC coupling) and strongly amplified. Sensitivity enhancement by several orders of magnitude was demonstrated by Kish and coworkers [29] in Taguchi sensors and by Ederth and coworkers [31] in nanoparticle films.

Significantly increased *selectivity* can be expected depending on the type of sensor and types of available FES "fingerprints". We define the selectivity enhancement by the factor specifying how many classical sensors a fluctuation-enhanced sensor can replace. When using power density spectra, the theoretical upper limit of selectivity enhancement is equal to the number of spectral lines. At typical experiments that is about 10000. However, when the elementary fluctuations are random-telegraph signals (RTSs) the underlying elementary spectra are Lorentzians [35,36] and the situation is less favorable because their spectra strongly overlap. As a consequence, experiments with COTS sensors indicate that the response of spectral lines against agent variations is often not independent. In a simple experimental demonstration with COTS sensors, a selectivity enhancement of six was easily reachable [18]. However, nanosensor development may be able to use all of the spectral lines more independently. Because both the FES signal in macroscopic sensors and the natural conductance fluctuations of the resistive sensors usually show $1/f$-like spectra [35,36], the lower the inherent $1/f$ noise strength in the sensor the cleaner the sensory signal. An interesting analysis can me made if we suppose that we shrink the sensor size so much that the different agents probe different RTS signals. Then principles for $1/f$ noise generation [37,38] indicate that one can resolve at most a few Lorentzian components in a frequency decade. Supposing six decades of frequency, the maximal selectivity enhancement would be around 18, supposing three fluctuators/decade.

With bispectra [21,26,27], the potential of selectivity increase is even greater because bispectra are two-dimensional data arrays. In the case of 10000 spectral lines, as mentioned above, the theoretical upper limit of selectivity increase is 100 million, but in the Lorentzian fluctuator limit that number is again radically reduced. Bispectra sense only the non-Gaussian part of the sensor signal, and for the utilization of the full advantages of bispectra it seems necessary to build the sensor within the submicron characteristic size range in order to utilize elementary microscopic switching events as non-Gaussian components. Moreover, the sixfold symmetry of the bispectrum function yields a further reduction of information by roughly a factor six. Using the above-mentioned estimation with three Lorentzian fluctuators/decade, over six decades of frequency, the selectivity enhancement would be around 50. It should be noted that this enhancement is independent from the spectral enhancement discussed above because bispectra probe the non-Gaussian components [35].

## 3. INFORMATION CHANNEL CAPACITY IN RESISTIVE GAS SENSORS

### 3.1. Information channel capacity in resistive gas sensors

Using Shannon's formula of information channel capacity in analog channels, it has recently been shown [35] that, in the case when the probing current density in the sensor is homogeneous and the sensor resistance fluctuations in the reference gas have 1/f spectrum, classical resistive sensors have the following upper limit of information flow rate:

$$C \leq \frac{1}{2t_m} \ln\left[1 + \frac{8p^2 V(R-R_0)^2}{AR^2}\right] = \frac{1}{2t_m} \ln\left[1 + \frac{8p^2 A_S d(R-R_0)^2}{AR^2}\right] \quad \text{(bit/second).} \tag{2}$$

Where $t_m$ is the measurement time window, $R$ and $R_0$ are the resistance response in the test gas and in the reference gas, respectively. $V$ is the volume of sensor film, $A_S$ is the surface of the sensor film and $d$ is its thickness. The strength of the 1/f noise in the film is characterized with the well known semiempirical formula:

$$\frac{S_u(f)}{U^2} = \frac{A}{Vf} \tag{3}$$

where $f$ is the frequency and A is the factor describing the strength of 1/f noise.

According to Eq. (2), in the practical (1/f-noise-dominated) limit and at fixed measurement time and film thickness, the larger the surface of the classical resistive sensor the greater the information channel capacity. However, in the limit of a sufficiently large agent concentration, the saturation time is controlled by the underlying diffusion processes taking place through the thickness of the film; therefore, in this case, the shortest measurement time is also controlled by diffusion according to

$$t_{m,min} \propto \left(\frac{d}{D}\right)^2, \tag{4}$$

where $D$ is the diffusion coefficient of the agent and/or its fragments through the film. *Therefore the thinner and larger the film the greater the information channel capacity.* This fact indicates that, in classical films, small thickness and large surface are preferable. From these two, the thickness is the dominant control parameter.

## 4. Information channel capacity in fluctuation-enhanced gas sensors

Suppose, the power density spectrum of the resistance fluctuations in a FES sensor has *K* different frequency ranges, in which the dependence of the response on the concentration of the chemical species is different from the response in the other ranges, one can write [17-19]:

$$\Delta S(f_1) = A_{1,1} C_1 + A_{1,2} C_2 + \ldots + A_{1,N} C_N$$
$$\vdots \tag{6}$$
$$\Delta S(f_K) = A_{K,1} C_1 + A_{K,2} C_2 + \ldots + A_{K,N} C_N$$

where $\Delta S(f_i)$ is the change of the power density spectrum of resistance fluctuations at the i-th characteristic frequency band, and the $A_{i,j}$ quantities are calibration constants in the linear response limit. Thus, a single sensor is able to provide a set of independent equations to determine the gas composition around the sensor. The number $K$ of different applicable frequency ranges has to be greater than or equal to the number $N$ of chemical species, i.e., $K \geq N$. This idealized situation needs sensors with linear response. Taguchi sensors do not produce linear response, see Figure 6, however future nanoscale devices may provide this property by the linear superposition of elementary fluctuations.

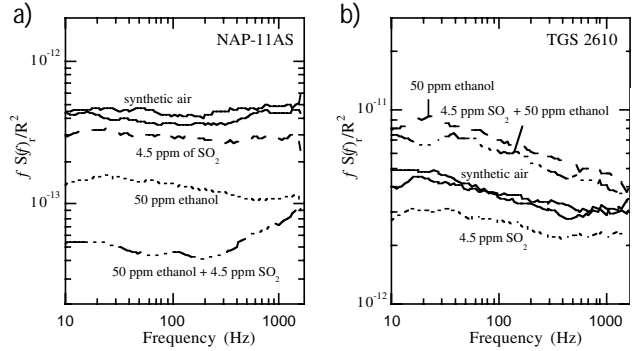

Figure 6. Experimental proof that the response in Taguchi sensors in not additive (nonlinear) [30].

In [35], the information channel capacity of power density spectrum based FES was estimated by assuming equal frequency bands and supposing that the relative error of the spectrum is much less than unity, then the information channel capacity (roughly) scales as:

$$C \propto \frac{t_m t_w f_s^2}{\mathrm{D}f} \qquad (12)$$

where $t_w$ is the duration (time window) of a single data sequence (for a single Fourier transformation), $t_m$ is the total measurement time (the elementary power spectra are averaged over that) and $f_s$ is the sampling frequency. It is supposed that the FES measurement starts after the sensor reached a stationary state in the test gas and that $t_m$ is much longer than the time needed to reach the stationary state, and that condition supposes thin sensor film just like in the classical sensor considerations above.

In such a case, the most important conclusion of Eq. 6 is that, in resistive FES applications, *the sensor surface can be very small without limiting the performance* (as large as enough Lorentzian fluctuators are present for each frequency band) because measurement time related statistical inaccuracies limit the information channel capacity, not a background noise.

In [35], similar conclusions are obtained for bispectrum based FES sensors.

## 5.  ON THE PRACTICAL RECOGNITION OF SPECTRA AND ROC CURVES

In the rest of the paper, as an example for the advantage of o\linear response, we show the result of simulations in a simple nanostructured FES sensor arrangement, which is based on one-dimensional diffusion noise [24].

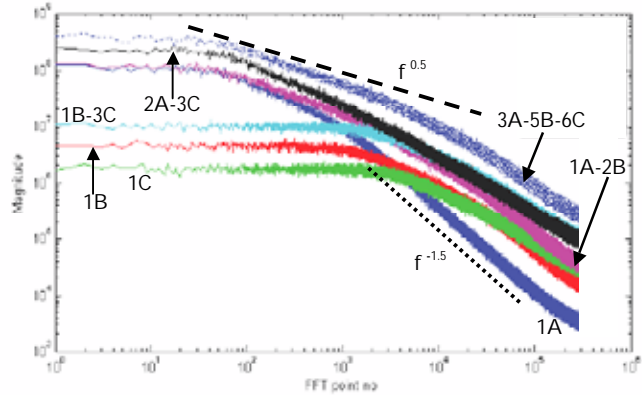

Figure 7. Spectra of the simulated situations. Molecules A, B ad C. Notations: 1A: 1 molecule A; 1B: 1 molecule C; 1A-2B: 1 molecule A and 2 molecules B; etc. The white noise at low frequencies is caused by the diffusion barriers.

Molecules A, and/or B, and/or C were executing a stationary diffusion (random walk), with molecule-specific diffusion coefficients, over the surface of the sensor with an active zone (sweet spot) at the middle and diffusion barriers at the side. The geometry of the system provides a one-dimensional diffusion noise [23-24], see Figure 7. Molecules over the sweet spot provide uniform output level, independently of the type of the molecule. In the low-concentration limit, the molecules move independently and the spectra are adding up linearly.

This system excellently shows the advantages of FES because the DC average signals of a single A, B or C molecule or that of equal concentrations of them are identical. Therefore, the classical sensor output can tell us only the total number of molecules on the surface (total molecular concentration) with zero selectivity/specificity.

However, FES can provide very much more information, especially, if proper analyzer and pattern recognizer tools are available. The results shown below were obtained by Signal Processing Co. (SPC) by using their hyperspectral analysis tool [36] which is a proprietary algorithm for the efficient determination of the linearly added spectral components (see Eqs. 6. Figures 8 and 9 show the actual and estimated abundance (molecule number) data. It is obvious that the algorithm performs much better then the naked eye when analyzing the spectra..

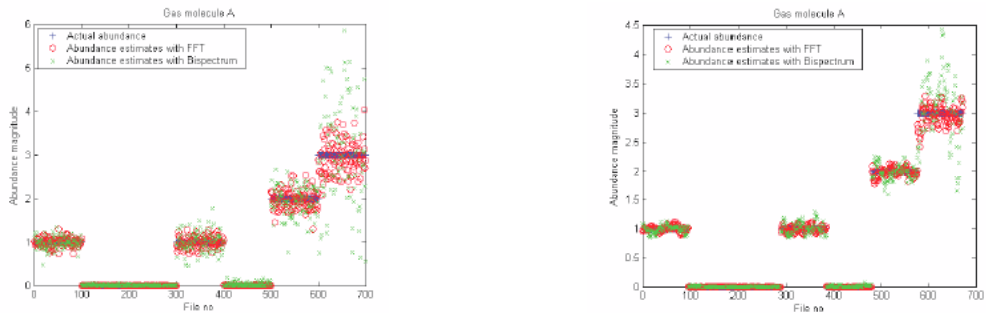

Figure 8. Abundance (molecule number or concentration) data for molecule A (see Figure 7). Actual data (+), Estimates with power density spectra (o) and estimates with bispectra (x). Left: 1 million data points/spectrum. Right: 5 million data points/spectrum). The X axis is the index number of data files; each actual abundance was tested 100 times (100 files each).

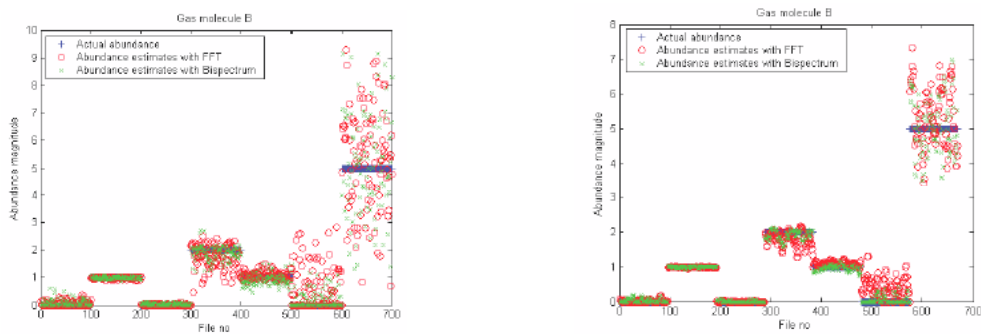

Figure 9. Abundance (molecule number or concentration) data for molecule B (see Figure 7). Actual data (+), Estimates with power density spectra (o) and estimates with bispectra (x). Left: 1 million data points/spectrum. Right: 5 million data points/spectrum). The X axis is the index number of data files; each actual abundance was tested 100 times (100 files each).

Receiver operating characteristic (ROC) curves are standard characterization tools of the quality of a sensing system. In Figure 10, the ROC curves generated by varied threshold values from the abundance estimations based on the simulations (see Figure 7). A threshold (quantization) value was defined to estimate if the abundance value is 0 or 1. Then this threshold value was varied so that the system went from the zero detection probability to the 1 detection probability and the detection and false alarm probabilities were recorded and plotted. For molecule A, power density and bispectra perform equally well, however for molecule B, the bispectra provide superior ROC curves.

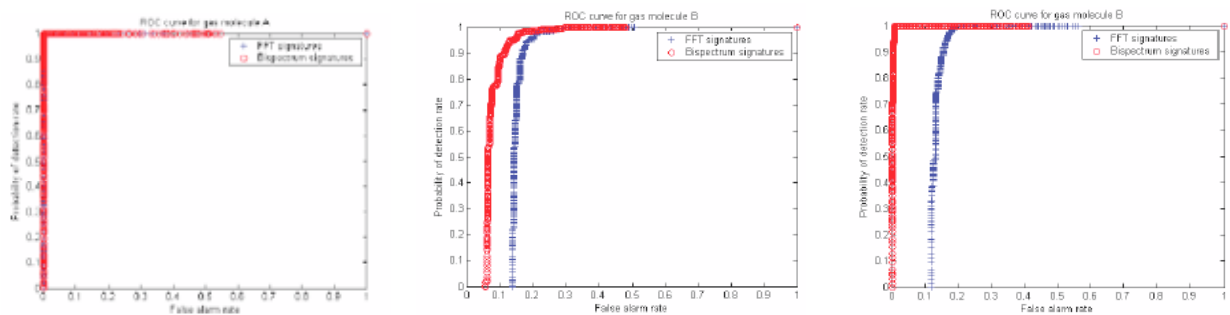

Figure 10. ROC curves generated by varied threshold values from the abundance estimations from the simulations (see Figure 7). For molecule A, power density and bispectra perform equally well, however for molecule B, the bispectra provide superior ROC curves.

## 6. CONCLUSION

Fluctuation enhanced gas sensing has very different requirements and superior characteristics compared to classical gas sensing. Further research is needed to develop sensors with linear response against the components in gas mixtures because these sensors provide the best FES performance.